\documentclass[11pt,notitlepage,prd,a4paper,
preprintnumbers,superscriptaddress,nofootinbib]{revtex4-1}

\usepackage[utf8]{inputenc}
\usepackage[colorlinks=true,citecolor=blue,linkcolor=blue]{hyperref}
\usepackage[normalem]{ulem}
\usepackage{url}
\usepackage{graphicx,wrapfig,float,slashed,cancel}
\usepackage{amsmath,amssymb,epsfig,graphicx,xcolor,stmaryrd}
\usepackage{bm}
\usepackage{enumitem}
\usepackage{multirow}

\definecolor{darkblue}{RGB}{1, 90, 173}

\def\nnb{\nonumber}

\begin{document}


\title{Re-analysis of rare radiative $\Xi_b^-\rightarrow \Xi^- \gamma$ decay in QCD}

\author{T.~M.~Aliev}
\email{ taliev@metu.edu.tr}
\affiliation{Physics Department, Middle East Technical University, 06531, Ankara, Turkey}
\author{A.~Ozpineci}
\email{ozpineci@metu.edu.tr}
\affiliation{Physics Department, Middle East Technical University, 06531, Ankara, Turkey}
\author{Y.~Sarac}
\email{ yasemin.sarac@atilim.edu.tr}
\affiliation{Electrical and Electronics Engineering Department,
Atilim University, 06836 Ankara, Turkey}

\date{\today}

\preprint{}

\begin{abstract}

The upper limit of the branching ratio of the  rare $\Xi_b^-\rightarrow \Xi^- \gamma$ decay is obtained as $BR(\Xi_b^-\rightarrow \Xi^- \gamma)<1.3\times10^{-4}$ by the LHCb. In the present work we study this decay within the light cone QCD sum rules employing the $\Xi_b$ distribution amplitudes. At first stage, the form factors entering the $\Xi_b^-\rightarrow \Xi^- \gamma$  decay are obtained. Next, using the results for the form factors the corresponding branching ratio for this decay is estimated to be $BR(\Xi_b^-\rightarrow \Xi^- \gamma)=(4.8\pm 1.3)\times 10^{-5}$. This value lies  below the upper limit established by the LHCb collaboration. Our finding for the branching ratio is also compared with the results of the other theoretical approaches existing in the literature.  
 
\end{abstract}


\maketitle

\renewcommand{\thefootnote}{\#\arabic{footnote}}
\setcounter{footnote}{0}
\section{\label{sec:level1}Introduction}\label{intro}

The exclusive weak decays of hadrons governed  by the flavor-changing neutral current (FCNC) $b \rightarrow s (d)$ transitions are forbidden in the Standard Model (SM) at the tree level and occur only at the one-loop level. Consequently, these decays hold exceptional significance for testing the predictions of the SM at the loop level as well as looking for the evidence of new physics beyond the SM. These decay channels are strongly suppressed, and this makes their experimental investigation difficult.

The rare exclusive radiative decay $\Xi_b^-\rightarrow \Xi^- \gamma$ induced by $b \rightarrow s$ transition has not been observed experimentally yet, and the LHCb collaboration imposed an upper limit on its branching ratio, $BR(\Xi_b^-\rightarrow \Xi^- \gamma)<1.3\times10^{-4}$~\cite{LHCb:2021hfz}. This decay was investigated  within different approaches, such as light-front quark model~\citep{Geng:2022xpn}, relativistic quark-diquark model~\cite{Davydov:2022glx}, light cone QCD sum rules~\cite{Olamaei:2021eyo,Liu:2011ema} using the $\Xi$ baryon distribution amplitudes, and in the framework of SU(3) flavor symmetry~\cite{Wang:2020wxn}. The difference in the predictions obtained in the Refs.~\cite{Olamaei:2021eyo,Wang:2020wxn,Geng:2022xpn,Davydov:2022glx}, which are below the experimental upper limit, and ~\cite{Liu:2011ema} and especially the difference between the predictions of Refs.~\cite{Olamaei:2021eyo} and \cite{Liu:2011ema} despite being obtained using the same framework and same distribution amplitudes (DA's), require a more careful analysis of this decay channel. Therefore in this work, we investigate the $\Xi_b^-\rightarrow \Xi^- \gamma$ decay in the framework of light cone QCD sum rules by using the DA's of the $\Xi_b$ heavy baryon. The light cone QCD sum rules method (LCSR)~\cite{Braun:1997kw} is an extension of the traditional QCD sum rules~\cite{Shifman:1978bx}, and one of the powerful approaches among nonperturbative methods that yields predictions consistent with the experimental observations. In the LCSR the operator product expansion (OPE) is conducted over the twist of the operators, rather than the dimension of the operators as in the traditional QCD sum rules.

The organization of the work is as follows. In the next section, the LCSR for the transition form factors responsible for the $\Xi_b^-\rightarrow \Xi^- \gamma$ decay are obtained by using the $\Xi_b$ light cone DA's.  Sec.~\ref{III} is devoted to the numerical analyses for the relevant form factors obtained in the previous section. Moreover, the corresponding branching ratio is attained using their numerical values. Discussions and our conclusion are presented in Sec.~\ref{IV}.

\section{Form factors for the $\Xi_b^-\rightarrow \Xi^- \gamma$ decay in light cone QCD sum rules }\label{II}

The rare $b\rightarrow s$ transition is described by the following effective Hamiltonian:
\begin{eqnarray}
\mathcal{H}^{\mathrm{eff}}=-\frac{G_F}{\sqrt{2}}V_{tb}V^{*}_{ts}\Big[\sum_{i=1}^{6}C_i(\mu)\mathcal{O}_i(\mu)+C_{7\gamma}(\mu)\mathcal{O}_{7\gamma}(\mu)+C_{8G}(\mu)\mathcal{O}_{8G}(\mu)\Big],
\end{eqnarray}
where $G_F$ is Fermi coupling constant, $C_i(\mu)$ are the Wilson coefficients, and $V_{tb}$ and  $V^{*}_{ts}$ are Cabibbo-Kobayashi-Maskawa matrix elements. The $\mathcal{O}_i$ are the local operators whose explicit forms can be found in Ref.~\cite{Davydov:2022glx}. Since the penguin operator $\mathcal{O}_{7\gamma}$ gives the main contribution to the $b\rightarrow s \gamma$ transition, the effective Hamiltonian for this transition is given as  
\begin{eqnarray}
\mathcal{H}^{\mathrm{eff}}=-\frac{G_Fe}{4\pi^2 \sqrt{2}}V_{tb}V^{*}_{ts} C_{7\gamma}^{\mathrm{eff}}(m_b)\bar{s}\sigma_{\mu\nu}\Big[m_b\frac{(1+\gamma_5)}{2}+m_s\frac{(1-\gamma_5)}{2}\Big]bF^{\mu\nu},
\end{eqnarray}
where we use $C_{7\gamma}^{\mathrm{eff}}(m_b)$ from Ref.~\cite{Davydov:2022glx}. The amplitude of the considered transition is obtained from the matrix element of the Hamiltonian taken between the initial and final states, which requires calculating the matrix element between the baryon states which can be expressed in terms of the form factors. In this section, we provide the details of the light cone QCD sum rule calculations to obtain the form factors for the $\Xi_b^-\rightarrow \Xi^- \gamma$ transition.

The basic object of the light cone QCD sum rules is the correlation function that sandwiches the time-ordered product of the interpolating current of the final baryon state and the weak transition current between the vacuum and the initial hadron state $\Xi_b$, i.e. 
\begin{equation}
\Pi_{\mu}(p,p')=i\int d^{4}xe^{ip'\cdot
x}\langle 0|\mathcal{T} \{J_{\Xi}(x)J_{\mu}(0)\}|\Xi_b(p,s)\rangle,
\label{eq:CorrF}
\end{equation}
where  $\mathcal{T}$ is the time ordering operator. For the considered problem the form of the weak transition current is $J_{\mu}=\bar{s}\sigma_{\mu\nu}(1+\gamma_5)q^{\nu}b$ and $ J_{\Xi}$ is the interpolating current of the $\Xi^-$ baryon
\begin{eqnarray}
J_{\Xi}=2\epsilon^{abc} \sum_{i=1}^{2}(s_a A_i d_b) B_i s_c \label{CurrentXi},
\end{eqnarray}
where $a,~b$ and $c$ are color indices, $A_1=C$, $A_2=C\gamma_5$, $B_1=\gamma_5$ and $B_2=\beta$ with $C$ representing the charge conjugation operator, $\beta$ is an arbitrary parameter.

In the light cone QCD sum rules method, the correlation function is calculated in terms of hadron and in terms of quark gluon degrees of freedom, respectively. After matching the results of both representations the desired sum rules for the physical quantities are obtained.

The hadronic representation of the correlation function is obtained by inserting a complete set of baryon states carrying the same quantum numbers as the interpolating current $J_{\Xi^-}$ in Eq.~(\ref{eq:CorrF}) and isolating the pole term of the $\Xi^-$ baryon we get 
\begin{eqnarray}
\Pi_{\mu}^{\mathrm{Had}}(p,p')= \frac{\langle 0|J_{\Xi}|\Xi(p',s')\rangle \langle \Xi(p',s')|J_{\mu}|\Xi_b(p,s)\rangle}{m_{\Xi}^2-p'^2}+\cdots.
\label{eq:Had}
\end{eqnarray}
The first matrix element appearing in Eq.~(\ref{eq:Had}) is determined in standard way and given as
\begin{eqnarray}
\langle 0|J_{\Xi}|\Xi(p',s')\rangle &=& \lambda u(p',s'),
\label{eq:matrixelement1}
\end{eqnarray}
where $\lambda$ and $u(p',s')$ represent the residue and spinor of the $\Xi^-$ baryon, respectively. The transition  matrix element, $\langle \Xi(p',s')|J_{\mu}|\Xi_b(p,s)\rangle$, is parametrized by the set of form factors in the following way:
\begin{eqnarray}
\langle \Xi(p',s')|\bar{s}i\sigma_{\mu\nu}q^{\nu}(1+\gamma_5)b|\Xi_b(p,s)\rangle &=& \bar{u}(p',s')\Big\{\frac{f_1^T}{m_{\Xi_b}}(\gamma_{\mu}q^2-\not\!q q_{\mu})+i f_2^T\sigma_{\mu\nu}q^{\nu}+ \frac{g_1^T}{m_{\Xi_b}}(\gamma_{\mu}q^2-\not\!q q_{\mu})\gamma_5 \nonumber\\
&+&ig_2^T \sigma_{\mu\nu}q^{\nu}\gamma_5\Big\}u_{\Xi_b}(p,s),
\label{eq:matrixelement2}
\end{eqnarray}
where $m_{\Xi_b}$ is the mass of the heavy $\Xi_b$ baryon.

In the considered problem the photon is real. Consequently, only two form factors, $f_2^T$ and $g_2^T$, at $q^2=0$ point contribute to the $\Xi_b^-\rightarrow\Xi\gamma$ decay. Therefore, in the next calculations, we concentrate on the computations of only the $f_2^T(0)$ and $g_2^T(0)$ form factors.

Substituting Eqs.~(\ref{eq:matrixelement1}) and~(\ref{eq:matrixelement2}) in the Eq.~(\ref{eq:Had}) and using the completeness relation $\sum u(p',s')\bar{u}(p',s')=\slashed{p'}+m_{\Xi}$ the correlation function for the hadronic side becomes
\begin{eqnarray}
\Pi_{\mu}^{\mathrm{Had}}(p,p')&=& \frac{\lambda}{m_{\Xi}^2-p'^2}\Bigg\{f_2^T\big[-(m_{\Xi_b} + m_{\Xi})q_{\mu}-2m_{\Xi_b} \not\!q v_{\mu}+(m_{\Xi_b}^2 - m_{\Xi}^2)\gamma_{\mu}+(m_{\Xi_b} + m_{\Xi})\not\!q\gamma_{\mu} +\not\!q q_{\mu} \big]+\nonumber\\
&+&g_2^T\big[(m_{\Xi_b} - m_{\Xi})q_{\mu}-2m_{\Xi_b} \not\!q v_{\mu}+(m_{\Xi_b}^2 - m_{\Xi}^2)\gamma_{\mu}-(m_{\Xi_b} - m_{\Xi})\not\!q \gamma_{\mu} +\not\!q q_{\mu} \big]\gamma_5\Bigg\}u_{\Xi_b}(p,s)\nonumber\\
&+&\cdots,
\label{eq:Had1}
\end{eqnarray}
in which $v_{\mu}$ is defined as $v_{\mu}=\frac{p_{\mu}}{m_{\Xi_b}} $.

The calculation of the correlation function for the QCD side proceeds as follows. Using the interpolating current for $\Xi^-$ baryon and the weak transition explicitly and after contracting the $s$-quark fields via Wick theorem, we obtain the correlation function as
\begin{eqnarray}
\Pi_{\mu}^{\mathrm{QCD}}(p,p')&=&i2\epsilon_{abc}\int d^{4}xe^{ip'\cdot x} \sum_{1}^{2}(A_i)_{\alpha\beta}(B_i)_{\rho\gamma}(i\sigma_{\mu\nu}q^{\nu}(1+\gamma_5))_{\sigma\zeta} \nonumber\\
&\times&\Bigg\{S_{\gamma\sigma}(x)\langle 0| s^{a}_{\alpha}(x)d^{b}_{\beta}(x)b^{c}_{\zeta}(0)|\Xi_b(p,s)\rangle 
+S_{\alpha\sigma}(x)\langle 0| s^{a}_{\gamma}(x)d^{b}_{\beta}(x)b^{c}_{\zeta}(0)|\Xi_b(p,s)\rangle\Bigg\},
\label{eq:QCD1}
\end{eqnarray}
where $S_{\alpha\sigma}(x)$ is the $s$-quark propagator. The matrix element in Eq.~(\ref{eq:QCD1}),  $\langle 0| s^{a}_{\alpha}(x)d^{b}_{\beta}(x)b^{c}_{\zeta}(0)|\Xi_b(p,s)\rangle $, is expressed in terms of the light-cone distribution amplitudes (DA's) of $\Xi_b^-$ baryon that have been studied in Ref.~\cite{Ali:2012zza}. Here we would like to note that the light-cone distribution amplitudes are obtained within the heavy quark effective theory. The relation between the heavy baryon state and the heavy baryon state in the heavy quark effective theory is given by $|\Xi_b(p)\rangle=\sqrt{m_{\Xi_b}}|\Xi_b(v)\rangle$. After these remarks, in Eq.~(\ref{eq:QCD1}) we make the replacement, $|\Xi_b(p)\rangle \rightarrow |\Xi_b(v)\rangle$, hence  appears the following matrix element 
\begin{eqnarray}
 \epsilon_{abc}\langle 0| s^{a}_{\alpha}(x)d^{b}_{\beta}(x)b^{c}_{\zeta}(0)|\Xi_b(v)\rangle .
\end{eqnarray}
This matrix element can be written in terms of $\Xi_b$ baryon DA's~\cite{Ali:2012zza} as
\begin{eqnarray}
\epsilon_{abc} \langle 0| s^{a}_{\alpha}(t_1n)d^{b}_{\beta}(t_2n)h^{c}_{\zeta}(0)|\Xi_b(v)\rangle =\sum_{j=1}^{4}a_j(\Gamma_j)_{\alpha\beta}u_{\zeta}(v),
\end{eqnarray}
where $h(0)$ is the heavy quark effective field coming from the replacement of heavy quark field $b(0)\rightarrow h(0)$, and
\begin{equation}
  \label{eq:DA}
  \begin{aligned}
    a_1 &= \frac{1}{8}f^{(2)} \psi_2(t_1,t_2)                & \Gamma_1 &= \bar{\slashed{n}}\gamma_5 C, \\
    a_2 &= -\frac{1}{8} f^{(1)} \psi_{3\sigma}(t_1,t_2)       & \Gamma_2 &= i \sigma_{\xi\varphi} \bar{n}^{\xi} n^{\varphi}\gamma_5 C, \\
    a_3 &= \frac{1}{4} f^{(1)} \psi_{3s}(t_1,t_2)             & \Gamma_3 &= \gamma_5C, \\
    a_4 &= \frac{1}{8}f^{(2)} \psi_{4}(t_1,t_2)               & \Gamma_4 &= \slashed{n}\gamma_5 C.
  \end{aligned}
\end{equation}
Here $\psi_2$, $\psi_{3 \sigma}(\psi_{3s} )$, and $\psi_4$ are the DA's with twist 2, 3, and 4, respectively. The light cone vectors $n$ and $\bar{n}$ have the following forms:
\begin{equation}
  \label{eq:n}
  \begin{split}
    n_{\alpha} &= \frac{1}{v x} x_{\alpha} \\
    \bar{n}_{\alpha} &= 2 v_{\alpha} - \frac{1}{v x} x_{\alpha} \\
    \bar{v}_\mu &= n_\alpha - v_\alpha,
  \end{split}
\end{equation}
and the DA's are defined as
\begin{equation}
  \label{eq:DA}
  \psi(t_1,t_2)=\int_0^\infty dw w\int_0^1 du e^{-iw (t_1 u+t_2 \bar{u})}\psi(u,w),
\end{equation}
with $\bar{u}=1-u$, $t_i=vx_i$ and $w$ being the total momentum of the light quarks.

Choosing the coefficients of the structures $\slashed{q}v_{\mu}$ and $\slashed{q} \gamma_5 v_{\mu}$ for the QCD part of the correlation function we have
\begin{eqnarray}
\Pi_{\mu}^{\mathrm{QCD}}(p,p')&=&\int du \int dw\Bigg\{\Bigg[\frac{1}{\Delta}[3 (1 + \beta) f^{(1)} \hat{\psi}_{3\sigma}(u,w) + (\beta-1) f^{(2)} m_s w \psi_2(u,w) \nonumber\\& +& (1 + 5 \beta) f^{(1)}(m_{\Xi_b} -  w) w \psi_{3s}(u,w) ]+\frac{1}{\Delta^2}(m_{\Xi_b} -  w) [  (\beta-1) f^{(2)} m_s  (\hat{\psi}_{2}(u,w) - \hat{\psi}_4(u,w)) \nonumber\\
&+& 2 (1 + \beta) f^{(1)} \hat{\psi}_{3\sigma}(u,w) q.v]\Bigg](\slashed{q} \gamma_5 v_{\mu}+\slashed{q} v_{\mu}) +\cdots\Bigg\},
\label{eq:PiQCD}
\end{eqnarray}
where $\cdots$ represent the contributions coming from other structures, the function $\hat{\psi}(u,w)$ is defined as
\begin{eqnarray}
\hat{\psi}(u,w) = \int_0^w d\tau \tau \psi(u,\tau),
\end{eqnarray}
 and
\begin{eqnarray}
\Delta = -m_s^2 - m_{\Xi_b} w + w^2 + p'^2 (1 - \frac{w}{m_{\Xi_b}}).
\end{eqnarray}

Matching the coefficients of the structures written in the above equation explicitly, $\slashed{q}v_{\mu}$ and $\slashed{q} \gamma_5 v_{\mu}$, obtained in both hadronic and QCD sides, and performing the Borel transformation with respect to the variable $p'^2$, we attain the following desired sum rules for the form factors $f_2^T(0)$ and $g_2^T(0)$:
\begin{eqnarray}
-2f_2^T(0) m_{\Xi_b} \lambda e^{-\frac{m_{\Xi}^2}{M^2}}&=&\Pi_1^B,\nonumber\\
-2g_2^T(0) m_{\Xi_b} \lambda e^{-\frac{m_{\Xi}^2}{M^2}}&=&\Pi_2^B.
\end{eqnarray}
where $\Pi_1^B$ and $\Pi_2^B$ represent the Borel transformed results obtained from the QCD side for the structures $\slashed{q}v_{\mu}$ and  $\slashed{q} \gamma_5 v_{\mu}$, respectively. From Eq.~(\ref{eq:PiQCD}) it follows that $\Pi_1^B=\Pi_2^B$, hence $f_2^T(0)=g_2^T(0)$. To obtain the results after Borel-transformation and continuum subtraction, we apply the master formula given as 
\begin{eqnarray}
\int_0^{\infty} dw\frac{\rho(u,w)}{\Delta^k}&=&(-1)^k\int_0^{w_0}dwe^{-\frac{s}{M^2}}\frac{\rho(u,w)}{(k-1)!(1-\frac{w}{m_{\Xi_b}})^{k}(M^2)^{k-1}}\nonumber\\
&-&\Bigg[\frac{(-1)^{k-1}}{(k-1)!}e^{-\frac{s}{M^2}}\sum_{j=1}^{k-1}\frac{1}{(M^2)^{k-j-1}}\frac{1}{s'}\Big(\frac{d}{dw}\frac{1}{s'}\Big)^{j-1}\frac{\rho(u,w)}{(1-\frac{w}{m_{\Xi_b}})^k}\Bigg]_{w=w_0},
\end{eqnarray}
where $s= \frac{m_s^2}{(1-\frac{w}{m_{\Xi_b}})} + w m_{\Xi_b}$, and $w_0$ is the solution of the equation $s=s_{0}$, where $s_{0}$ is the continuum threshold.

Using the matrix element given in the Eq.(\ref{eq:matrixelement2}) the decay width of the rare  $\Xi_b^-\rightarrow \Xi^-\gamma$ radiative decay is
\begin{equation}
\Gamma=\frac{G_F^2\alpha_{em}}{64\pi^4}|V_{tb}V_{ts}^{*}|^2 m_b^2 |C_{7\gamma}^{(0)eff}(m_b)|^2\Big(\frac{m_{\Xi_b}^2-m_{\Xi}^2}{m_{\Xi_b}}\Big)^3\Big[(1+\frac{m_s}{m_b})^2|f_2^T(0)|^2+(1-\frac{m_s}{m_b})^2|g_2^T(0)|^2\Big],
\label{eq:DW}
\end{equation}
where $C_{7\gamma}^{(0)eff}(m_b)=-0.310$~\cite{Davydov:2022glx,Buras:1993xp}, $G_F = (1.166 \times 10^{-5})~\mathrm{GeV}^{-2}$ and the fine structure constant is $\alpha_{em}\equiv \frac{e^2}{4\pi}=\frac{1}{137}$.

\section{Numerical Analyses}\label{III}

In the previous section, the sum rules for the form factors $f_2^T$ and $g_2^T$ at $q^2=0$ point are derived. In present section, we perform the numerical analyses of sum rules for the form factors. Moreover, using the obtained results for $f_2^T(0)$ and $g_2^T(0)$ we estimate the corresponding branching ratio.

The main input parameters for the LCSR are the distribution amplitudes (DA's), which are the DA's of $\Xi_b$ baryon in our case. These DA's were obtained in Ref.~\cite{Ali:2012zza}, and their expressions are
\begin{eqnarray}
\label{eq:DAsfunc}  
\psi_2(u,w) &=& w^2 \bar{u}u \sum_{n=0}^2 {a_n\over \varepsilon_n^4}
{C_n^{3/2} (2 u -1) \over | C_n^{3/2} |^2 } e^{-w/\varepsilon_n}~, \nnb \\
\psi_4(u,w) &=&  \sum_{n=0}^2 {a_n\over \varepsilon_n^2}
{C_n^{1/2} (2 u -1) \over | C_n^{1/2} |^2 } e^{-w/\varepsilon_n}~, \nnb \\
\psi_3^{(\sigma,s)}(u,w) &=& {w \over 2}  \sum_{n=0}^2 {a_n\over
\varepsilon_n^3}
{C_n^{1/2} (2 u -1) \over | C_n^{1/2} |^2 } e^{-w/\varepsilon_n}~,
\end{eqnarray}
for which the values of the parameters $a_0,~a_1,~a_2$, and $\varepsilon_0,~\varepsilon_1,~\varepsilon_2$ are given in Ref.~\cite{Ali:2012pn} with $A=\frac{1}{2}$, $C_n^{\lambda}(2 u-1)$ is the Gegenbauer polynomial, and
\begin{eqnarray}
|C_n^{\lambda}|^2=\int_0^1 du [C_n^\lambda(2u-1)]^2.
\end{eqnarray}

The values of the other input parameters are as follows: For the residue $\lambda$, we have used the result of Ref.~\cite{Lee:2002jb} obtained for $\tilde{\lambda}^2$ with $\tilde{\lambda} = (2\pi)^2\lambda$. In our analysis to get numerical values for $\lambda$ we have used the numerical values of the condensates given in Ref.~\cite{Lee:2002jb}  and the threshold and Borel parameters are varied in the ranges $2.5~\mathrm{GeV}^2 \leq s'_0 \leq 2.8~\mathrm{GeV}^2$ and $1.0~\mathrm{GeV}^2 \leq M'^2\leq 1.5~\mathrm{GeV}^2$, respectively. The parameters $f^{(1)}$ and $f^{(2)}$ are taken as $f^{(1)}=f^{(2)}=(2.23 \pm 0.35)\times 10^{-2}~\mathrm{GeV}^3$~\cite{Wang:2020mxk}. The input parameters taken from PDG~\cite{Workman:2022ynf} are $|V_{tb}|=1.014\pm 0.029$, $|V_{ts}|=(41.5\pm 0.9)\times 10^{-3}$, $m_b = 4.78\pm 0.06$~GeV, $m_s=93.4^{+8.6}_{-3.4}$~MeV, $m_{\Xi_b^-}=(5797.0\pm 0.6)$~MeV, $m_{\Xi^-}=(1321.71 \pm 0.07)$~MeV, $\tau_{\Xi_b^-}=(1.572\pm 0.040)\times 10^{-12}$~s.

The sum rules also contain following auxiliary parameters: Borel parameter, $M^2$, threshold parameter $s_0$ and the parameter $\beta$ entering to the interpolating current. The continuum threshold $s_0$ is determined from the analyses of two-point QCD sum rules, namely its value is obtained from the condition that the mass sum rule reproduces the experimentally measured values within 10\%  accuracy. This analysis leads to the result $2.5~\mathrm{GeV}^2 \leq s_0 \leq 2.8~\mathrm{GeV}^2$. The working region of the $M^2$ is determined by demanding that the power corrections and the continuum contributions be suppressed compared to the leading twist-2 contribution. Taking these conditions into account, we obtain the following domain for this parameter: $1.7~\mathrm{GeV}^2 \leq M^2\leq 2.5~\mathrm{GeV}^2$.

\begin{figure}[h!]
\begin{center}
\includegraphics[totalheight=6cm,width=9cm]{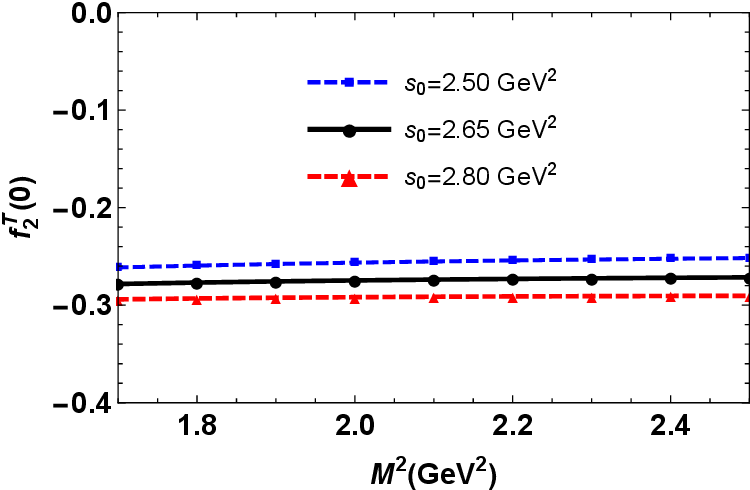}
\end{center}
\caption{ Variation of the the form factor $f_2^T(0)$ as function of $M^2$ at different values of threshold parameter $s_0$ and $\beta=-1$.}
\label{gr:FormfacMsq}
\end{figure}

\begin{figure}[h!]
\begin{center}
\includegraphics[totalheight=6cm,width=9cm]{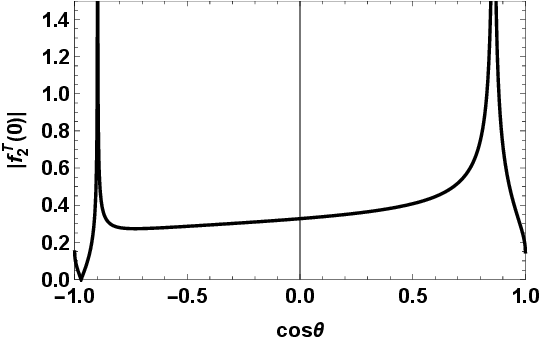}
\end{center}
\caption{ Variation of the the form factor $|f_2^T(0)|$ as function of $\cos\theta$ at fixed values of threshold parameter $s_0$ and Borel parameter $M^2$ in their working regions.}
\label{gr:Formfaccostheta}
\end{figure}

Using the DA's for $\Xi_b$ baryon given in Eq.~(\ref{eq:DAsfunc}), in Figure~\ref{gr:FormfacMsq} the dependence of the form factor, $f_2^T(0)$, at zero momentum transfer squared on $M^2$ at fixed values of $s_0$ and $\beta=-1$, is presented. We see that the form factor, $f_2^T(0)$, exhibits good stability when $M^2$ varies in the working region, as can be seen.
In order to find the working region of $\beta$, in Figure~\ref{gr:Formfaccostheta} we present the dependence of $f_2^T(0)$ on $\cos\theta$, where $\tan\theta=\beta$, at fixed values of $M^2$ and $s_{0}$ from their working regions. From this figure we observe that, when $\cos\theta$ varies in the region $-0.8\leq\cos\theta\leq 0.5$, the form factor $f_2^T(0)$ exhibits good stability on the variation of $\cos\theta$. Besides in this region the required criteria for Borel parameter, $M^2$, and threshold parameter $s_0$ including the convergence of the OPE, are satisfied. 

\begin{figure}[h!]
\begin{center}
\includegraphics[totalheight=6cm,width=9cm]{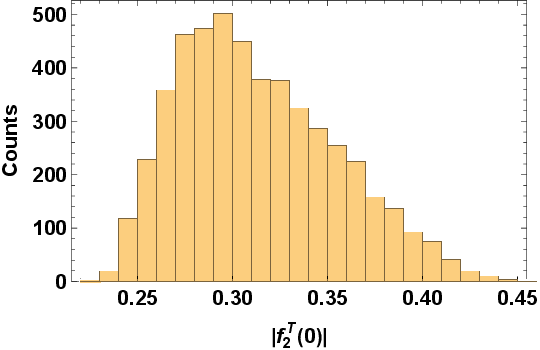}
\end{center}
\caption{ The histogram of the form factor $f_2^T(0)$ obtained using arbitrary values of the auxiliary parameters, $s_0$, $s'_0$, $M^2$, $M'^2$ and $\cos\theta$ from their working intervals.}
\label{gr:Hist}
\end{figure}

To analyze the stability of our predictions on all of the determined parameter space, the parameters, $s_0$, $s'_0$, $M^2$, $M'^2$ and $\cos\theta$, are randomly selected inside the chosen region. The histogram of 5000 such computations are shown in Figure~\ref{gr:Hist}.
 From these data the mean and the standard deviation of our predictions on the form factors are
\begin{eqnarray}
|f_2^T(0)|=|g_2^T(0)|= 0.31\pm 0.04.
\end{eqnarray}
Note that from the two-point sum rules results of Ref.~\cite{Lee:2002jb}, the sign of $\lambda$ can not be predicted. Hence we only show the absolute values of the form factors at $q^2=0$. Note also that the relative size of standard deviation is a measure of the stability of our predictions within the chosen region in the parameter space. 

After the determination of the form factors, $f_2^T(0)$ and $g_2^T(0)$, we can determine the decay width applying Eq.~(\ref{eq:DW}). Using the lifetime for $\Xi_b^-$ baryon, $\tau_{\Xi_b^-}=(1.572\pm 0.040)\times 10^{-12}$~s, we get the branching ratio as
\begin{eqnarray}
BR(\Xi_b^-\rightarrow\Xi^-\gamma)&=&  (4.8\pm 1.3)\times 10^{-5}. 
\end{eqnarray}

At the end of this section we compare our result on branching ratio of $\Xi_b^-\rightarrow\Xi^-\gamma$ with the existing results in literature and with the experimental upper bound, $BR(\Xi_b^-\rightarrow\Xi^-\gamma)<1.3\times10^{-4}$~\cite{LHCb:2021hfz}. The results are presented in the Table~\ref{table1}.
\begin{table}[]
\begin{tabular}{c|c}
\hline
References                   & The branching ratio values \\ \hline\hline
This work                                        &  $ (4.8\pm 1.3)\times 10^{-5} $             \\
Experiment~\cite{LHCb:2021hfz}                   &  $<1.3\times10^{-4} $                         \\
Light front quark model~\cite{Geng:2022xpn}      &  $ (1.1\pm 0.1)\times 10^{-5} $               \\
Relativistic quark model~\cite{Davydov:2022glx}  &  $(0.95 \pm 0.15) \times 10^{-5} $            \\ 
Light cone QCD sum rules~\cite{Olamaei:2021eyo}  &  $ (1.08^{+0.63}_{-0.49})\times 10^{-5} $     \\
Light cone sum rules~\cite{Liu:2011ema}          &  $ (3.03\pm 0.1)\times 10^{-4}$               \\
SU(3) flavor symmetry~\cite{Wang:2020wxn}        &  $ (1.23\pm 0.64)\times 10^{-5}  $            \\ \hline
\end{tabular}
\caption{The branching ratio, (BR), for $\Xi_b^-\rightarrow\Xi^-\gamma$ obtained in different frameworks and the experimental upper bound.}
\label{table1}
\end{table}
From the table, it follows that the result of Ref.~\cite{Liu:2011ema} exceeds the predictions of all other works by one order  and even exceeds the experimental upper limit. Our result has consistent order of magnitude with the results of all the works, except that of Ref.~\cite{Liu:2011ema}. The measurement of the branching ratio of $\Xi_b^-\rightarrow\Xi^-\gamma$ decay may be useful for distinguishing the right picture.

\section{Summary and conclusion}\label{IV}

By using the heavy $\Xi_b$ baryon distribution amplitudes the rare radiative $\Xi_b^-\rightarrow\Xi^-\gamma$  decay is studied within the light cone QCD sum rules. The sum rules for the relevant form factors are derived and their numerical values are determined at $q^2=0$ point. Using the results of the form factors the branching ratio is estimated. Moreover, we perform a comparison between our finding and the results of other works in the literature on the branching ratio of $\Xi_b^-\rightarrow\Xi^-\gamma$ decay. We obtained that the branching ratio, $BR(\Xi_b^-\rightarrow\Xi^-\gamma)=(4.8\pm 1.3)\times 10^{-5}$, has consistent order of magnitude with  results given in Refs.~\cite{Geng:2022xpn,Davydov:2022glx,Olamaei:2021eyo,Wang:2020wxn} and below the experimental upper limit~\cite{LHCb:2021hfz}. Besides, it is smaller than the result given in Ref.~\cite{Liu:2011ema}. Our final remark to this work is that the results presented here can be improved by taking into account $\mathcal{O}(\alpha_s)$ corrections to the distribution amplitudes, as well as improving the values of parameters appearing in them.




\end{document}